\documentstyle[editedvolume,numreferences,graphicx,floatflt]{crckapb}
\newcommand{\stt}{\small\tt}
\begin{opening}

\title{Astronomy Librarian -- Quo Vadis?}
\author{Jill Lagerstrom }
\institute{Space Telescope Science Institute\\
           3700 San Martin Drive\\
           Baltimore MD 21218, USA\\
           \stt lagerstrom@stsci.edu}
\author{Uta Grothkopf}
\institute{European Southern Observatory\\
           Karl-Schwarzschild-Stra{\ss}e 2\\
           D-85748 Garching, Germany\\
           \stt esolib@eso.org}

\end{opening}

\begin{document}
\setcounter{page}{111}

\begin{abstract}
``You don't look like a librarian" is a phrase we often hear in the astronomy department or observatory 
library. Astronomy librarians are a breed apart, and are taking on new and non-traditional roles as 
information technology evolves. This talk will explore the future of librarians and librarianship through 
the lens of some of the recent talks given at the sixth ``Libraries and Information Services in Astronomy"
conference held in Pune, India in February 2010. We will explore the librarian's universe, illustrating how 
librarians use new technologies to perform such tasks as bibliometrics, how we are re-fashioning our library 
spaces in an increasingly digital world and how we are confronting the brave new world of Open Access, to 
name but a few topics.
\end{abstract}

\section{Introduction}

``You don't look like a librarian" is a phrase we hear so often that Ruth Kneale, systems librarian at 
the National Solar Observatory, decided to write a book with this title (Kneale 2009), after so many years 
of blogging on the issues surrounding this phrase with frequent 
regularity\footnote{\tt http://desertlibrarian.blogspot.com}.  Our popular culture is 
fascinated by the image of the librarian. An advertisement for a Sony ebook reader states ``Sexier than a 
librarian." Noah Wyle plays a librarian who goes on several Indiana-Jones-type adventures in a series 
of made-for-TV-movies. And, last but not least, a long time ago in a galaxy far far away, there is even 
a Jedi librarian with whom Obi Wan Kenobi himself has a mild altercation over the existence of some 
astronomical data in the archive. Unfortunately, many of these portrayals, paradoxically even the 
``futuristic" ones, make librarians seem out-of-touch with modern technology and modern culture.

These entertaining portrayals aside, real librarians in astronomy libraries in all corners of the world 
are doing innovative and critical work to support the missions of the organizations they serve. To explore 
them, we will take a look at astronomy librarianship through the lens of some of the issues and projects 
discussed at the recent ``Libraries and Information Services in Astronomy (LISA) VI" conference held in 
Pune, India at the Inter-University Centre for Astronomy and Astrophysics in February 2010.  Excellent 
histories of the LISA conferences by Brenda Corbin and Uta Grothkopf may be found in Volume 7 of the
Organizations and Strategies in Astronomy series and in the proceedings of the fifth LISA 
conference\footnote{\tt http://www.eso.org/sci/libraries/lisa.html}  (Corbin 2006 \& Grothkopf 2007). 
Videos of many of the talks may be seen on the LISA VI 
website\footnote{\tt http://libibm.iucaa.ernet.in/conf/index.php/LISA/conf}.  
The proceedings of the conference will 
be published by the Astronomical Society of the Pacific Conference Series. Finally, we will conclude with 
further thoughts and observations about the image of librarians today and what it may mean for the future.

\section{The LISA VI Conference}

Astronomy librarians, appropriately, are a very international and diverse crowd. More than 90 participants 
from more than 18 countries attended LISA VI.  The theme of the conference was ``21st Century Astronomy 
Librarianship: From New Ideas to Action" and the six major topics discussed were: 
{\em The Future of Librarianship}, {\em Metrics}, {\em Open Access}, {\em Data Curation and Preservation}, 
{\em Virtual Communities} and {\em Use and Access}. While taking the reader on a tour of the highlights of 
the ``official" themes, we will explore some of the subcurrents running through the presentations that 
describe the constant adaptation and evolution of astronomy librarians and the diverse ways they serve 
their local communities. These examples, in no way, represent all that librarians do, but rather a subset 
of interesting and successful projects.

\subsection{The Future of Librarianship}

Discussions on the future of librarianship focus on the new and emerging roles that librarians play within 
the knowledge spheres of their organizations. In addition, they grapple with the role of the print collection 
in an increasingly electronic environment. Managing both print and online collections simultaneously with a 
matrix of heterogeneous user expectations and the re-envisioning of the library as a place are two 
subcurrents that run throughout these discussions. 

Advice on managing change, both practical and psychological, was offered by Jane Holmquist, librarian at 
Princeton University. Departmental science library collections at Princeton have been merged and housed 
in a new state-of-the-art building designed by Frank Gehry. Jane's situation is becoming more prevalent as 
universities centralize services. Librarians from the Universities of Oslo and Helsinki are also managing 
change brought about by similar consolidations within their local academic cultures.

Jill Lagerstrom's poster on ``If it's not on the web it must not exist" used the opportunity of a major 
renovation at the Space Telescope Science Institute to find out how much of the print collection is 
available online, and how much is indexed by the ADS. 

The frontier of ``E-science" is being explored by librarians. Lee Pedersen of Brown University chronicled 
her conversations with faculty about the storage, dissemination, manipulation, protection and publication 
of the data generated by their research.

Librarians are increasingly involved in ``Communicating astronomy with the public." Francesca Brunetti, of 
the INAF Osservatorio Astrofisico di Arcetri, argued that librarians provide the right mix of skills for 
this task: scientific authority and accuracy to the public, and the ability to deal with people of 
different ages, skills and heterogeneous educational backgrounds. 

\subsection{Metrics}

Librarians use publication statistics to analyze the scientific productivity and impact of their 
organizations, scientific instruments (such as telescopes), groups of people, topics/subjects or even 
individual scientists. The managers of the organizations that we serve need this information for 
decision-making purposes and to show funding agencies that money has been put to good use. Fortunately, 
methods have been developed to construct these metrics efficiently and wisely, and librarians are well 
versed in their practice. Discussion surrounding this topic deals not only with metrics projects and 
methods, but also with the shortcomings of information products (journal websites, ADS, Web of Science) 
when it comes to performing tasks such as these, which need to be done as accurately and systematically as 
possible.

Christopher Erdmann at the ESO Library has developed two software tools, FUSE and telbib, to manage the 
workflow specific to constructing bibliographies for citation analysis.  FUSE is a full-text searching 
utility which has features more sophisticated than any publisher's website. Telbib is a content management 
system that generates next-generation bibliometric statistics.

Eva Isaksson looked at how departmental mergers at the University of Helsinki have affected productivity and 
impact statistics using the ISI citation database. When an astronomy department merges with a physics 
department, does this affect the measure of the department's overall performance?

Nishtha Anilkumar examined the types of research used by doctoral students in their theses at the Physical 
Research Laboratory in Ahmedabad, India. Longitudinal statistics on the use of Open Access materials, 
non-subscribed journals, and e-prints paint a revealing portrait of the research landscape and how it 
evolves over time. This type of analysis can reveal gaps in the library's collections.

\subsection{Open Access}

The Open Access movement promotes the idea that scholarly works (most usually journal articles) should be 
freely accessible to everyone. Librarians have, in general, been vocal advocates of this idea, which comes 
in a variety of economic models and paradigms, because we are the ones who get the bills for the journals 
and often are forced to cancel access to journals as our budgets shrink. According to {\em Library Journal's}
periodicals price survey\footnote{\tt http://www.libraryjournal.com/article/CA6547086.html},  the average 
cost of an astronomy journal has been increasing at a rate of 8 to 10\% each year, certainly far higher than 
the rate of inflation. Astronomers have historically been early adopters of many aspects of Open Access: 
they deposit e-prints in astro-ph and the major society journals make articles freely available after two 
years. They are also accustomed to page charges, whereby the author shares the cost of publishing. 

To further complicate matters, the U.S. Office of Science and Technology Policy has recommended that the 
version of record of federally funded research be made freely available within one year of publication. 
Terry Mahoney, scientific editor at the Instituto de Astrof{\'\i}sica de Canarias, asked the question 
``Is Open Access Right for Astronomy?"  Of course, the answer is neither a simple ``yes" or ``no."
Terry took a refreshingly critical look at the economic realities of Open Access in the complex and 
heterogeneous world of astronomical publishing. Joe Jensen, managing editor of the Astronomical Society 
of the Pacific Conference Series also goes beyond the rhetoric to spell out what Open Access means 
specifically for society conference proceedings. Terry and Joe both reminded the audience that 
``there's no such thing as a free lunch" and that there is no one-size-fits-all solution as scholarly 
publishers confront Open Access.

For M.N. Nagaraj at the Raman Research Institute Library, Open Access is not just about journals. 
They have created an open digital repository of physics and astronomy theses created at their institution. 
Before the advent of electronic communication, theses, in particular, were often prohibitively difficult to 
obtain. A free repository is most welcome. However, these digital repositories don't just ``create 
themselves" and bring with them a host of complex issues, such as determining the best scanning methods, 
getting permissions and establishing metadata standards. 

\subsection{Data Curation and Preservation}

The curation and preservation of data raise many unanswered questions.  What is the shelf life of digital 
data? What does it cost to make data free? How can we cooperate and collaborate better to answer these 
questions? 

Copyright is perhaps the most complex issue for archivists who want to make the electronic documents in 
their collections freely available and use the web to promote their institutions. Christina Birdie at the 
Indian Institute of Astrophysics is confronting the challenge of balancing the need to disseminate 
information freely with the rights of authors under the law. Unpublished works are particularly difficult 
to manage.

According to Andras Holl, librarians at small observatories must be involved in the team responsible for 
constructing, managing and ensuring access to local data archives. He outlines a plan with guidelines for 
the Konkoly Observatory to establish such a data archive.

\subsection{Virtual Communities}

Virtual tools facilitate communication and collaboration across global networks. Librarians make use of 
these tools to build communities.

Leila Fernandez participates in a virtual community to engage the public with astronomy at the University 
of Toronto. An interactive chat facility combined with live viewing of telescope images enable a virtual 
community to participate in astronomical activities to which they would not normally have access. 
Participants from the southern hemisphere are able to view portions of the sky not normally visible to 
them. Groups such as bilingual people and First Nations Communities are targeted participants in this 
outreach program.

Social networking is best known to us as Facebook and Twitter. However, a host of book social network 
sites, such as LibraryThing and Shelfari, have also emerged. Francesca Martines, of the INAF Osservatorio 
Astronomico di Palermo, evaluated these with a librarian's bibliographic eye and contemplated their use 
in the astronomy library. Librarians often evangelize interesting new technologies to their local users. 

Alberto Accommazzi, ADS Project Manager, sees the ADS as a sort of virtual community, a collection of 
heterogeneous knowledge products interlinked with the ADS as a node. In order to take advantage of powerful 
web architectures, though, we need to structure our data accordingly. Alberto calls on us to do this so 
that we can fully collaborate, share and uncover networks of communication.

\subsection{Use and Access}

Librarians study their communities and attempt to understand their local users' specific needs. A good 
librarian is familiar with the culture not just of the specialized field she supports, but of her local 
institution as well. She is also familiar with the marketplace of new and emerging information products. 
Librarians seek the best products for their users with the intent of using library funds effectively and 
saving the user time.

Electronic books are an example where librarians are waiting for the future to happen. Molly White, at 
the University of Texas at Austin, surveyed her astronomy faculty to reveal their attitudes toward the 
ebooks available to them through the library. When asked if they prefer ebooks to print books, the 
responses were mixed. The answer was a resounding ``sort of."

Open source library catalogs have entered the marketplace. Uta Grothkopf, of the ESO Library, has become 
frustrated because commercial proprietary library catalog software is constantly behind the technological 
times. Investigating open-source alternatives gives her greater flexibility and allows her to configure 
these tools to provide her users with twenty-first century research capabilities. Uta evaluated the features 
of various open source catalogs and shared her experience with her endeavour to adopt an open source product 
in a small library.

Hemant Kumar Sahu of the Inter-University Centre for Astronomy and Astrophysics researched the specific 
information-seeking behaviors of astronomers and astrophysicists by using a questionnaire to learn about 
their use of books, online catalogs, print materials and online journals. With the gathered statistics, 
Sahu can paint a picture of the impact of Information and Communication Technology on the information-seeking 
behavior of Indian astronomers and astrophysicists. This picture can be used to inform strategic planning 
initiatives for the library's infrastructure.

\section{Conclusions}

\subsection{Subcurrents}

While there were six official themes to the talks at the LISA VI conference, many subcurrents can be found 
throughout such as: 
\begin{itemize}
\item
Playing active roles in supporting the communication of astronomy to the public
\item
Demonstrating the value of the intellectual output of their organizations through bibliometrics
\item
Pro-actively making local knowledge products freely available
\item
Becoming more user-centered and less library-centered
\item
Refocusing on the library as a place
\item
Exploring the frontier of e-science and data archives
\end{itemize}

\subsection{What's in a name?}

Will librarians remain librarians? Last year, our trade organization, the Special Libraries Association, 
voted on whether to change its name to the ``Association of Strategic Knowledge Professionals." The 
discussion surrounding this topic was very charged, indeed. The leadership of the organization was 
concerned that the term ``librarian" was a liability. While there were many members who were not necessarily 
attached to the name, which has been in use for 100 years, the result was that the name change was voted 
down.  The word ``libraries" has received some attention in the educational sphere as well. One-third of 
library schools have become ``iSchools."  A new type of knowledge worker that has grown out of librarianship 
is called the ``informationist." These new professionals, who primarily provide research assistance, came 
into existence in the context of clinical care and biomedical research. Informationists receive specialized 
training in their subject areas that goes beyond ``on-the-job" learning. Programs in informatics are popping 
up in universities with library curricula and have expanded beyond the biomedical to include fields such as 
chemistry and even music. ``Embedded" librarianship describes a set of practices librarians are adopting to 
get out of their offices and out from behind their desks.  Since more patrons are using online information, 
we do not see them as often and this is an effective way of making sure our services are visible. As 
librarians' roles evolve, along with our names in some cases, our traditional values of preservation and 
access to information remain.

\subsection{This Book is Overdue}

Marilyn Johnson, in her recent book {\em This Book is Overdue! How Librarians and Cybrarians Can Save Us All}
(Johnson 2010) makes the case that, despite ongoing technological advances, there will always be a need for 
human help. Librarians must continue to be advocates for their users, or, as a reviewer of this book puts 
it, ``pragmatic idealists who fuse the tools of the digital age with their love for the written word and 
the enduring values of free speech, open access, and scout-badge-quality assistance to anyone in 
need\footnote{\tt http://www.harpercollins.com/books/9780061431609/This\_Book\_Is\_Overdue/index.aspx}."

\subsection{Where are we going?}

The title of this paper asks of the astronomy librarian {\em ``Quo Vadis?"} which in Latin means 
``Where are you going?"  The answer, that we hope has become apparent from the many subcurrents running 
through the talks at the LISA VI conference, is ``we go where the information goes" and ``we go where our 
users go."  Our good faith, however, may not be enough. Terry Mahoney's Declaration concerning the evolving 
role of libraries in research centres warns us of the ``increasing invisibility of research libraries 
vis-\`a-vis recent accelerated changes in publishing and reader-access technology" and calls on librarians 
to ``adopt a more proactive stance in making their contribution known to the research communities they 
serve." (Mahoney 2007) Hopefully, recent efforts to combat the image of the librarian as antiquated and 
out-of-touch will help us to demonstrate that we are vital parts of the organizations we serve.


\begin{thebibliography}{99}

\bibitem{}
Corbin, B.G. \& Grothkopf, U. 2006, LISA -- The Library and Information Services in Astronomy conferences, 
in: {\em Organizations and Strategies in Astronomy (OSA)}, Vol. 7, A. Heck (ed.), Springer, Dordrecht, 
285-306. 
\bibitem{}
Grothkopf, U \& Corbin, B.G. 2007, LISA -- How We Got Where We Are Now (Closing Remarks) in: 
{\em Library and Information Services in Astronomy V}, S. Ricketts, C. Birdie E. Isaksson and C. Birdie 
(eds.), Astronomical Society of the Pacific, San Francisco, ASP Conference Series, 377, 25-26.
\bibitem{}
Johnson, M. 2010, {\em This book is overdue! : how librarians and cybrarians can save us all}, Harper, 
New York.
\bibitem{}
Kneale, R. 2009, {\em You don't look like a librarian : shattering stereotypes and creating positive new 
images in the Internet age}, Information Today, Inc., Medford, NJ.
\bibitem{}
Mahoney, T.J. 2007, Declaration concerning the evolving role of libraries in research centres, 
{\em The Observatory}, 127, 401-402.
\end{thebibliography}
\end{document}